\documentclass[preprint,aps,showpacs]{revtex4}
\usepackage{epsfig}
\begin{document}
\title{Multiple time scales and the exponential Ornstein-Uhlenbeck stochastic
volatility model}
\author{Jaume Masoliver and Josep Perell\'o}
\affiliation{Departament de F\'{\i}sica Fonamental, Universitat de Barcelona,\\
Diagonal, 647, E-08028 Barcelona, Spain}
\date{\today}

\begin{abstract}
We study the exponential Ornstein-Uhlenbeck stochastic volatility model and observe that the model shows a multiscale behavior in the volatility autocorrelation. It also exhibits a leverage correlation and a probability profile for the stationary volatility which are consistent with market observations. All these features make the model quite appealing since it appears to be more complete than other stochastic volatility models also based on a two-dimensional diffusion. We finally present an approximate solution for the return probability density designed to capture the kurtosis and skewness effects.
\end{abstract}
\pacs{89.65.Gh, 02.50.Ey, 05.40.Jc, 05.45.Tp}
\maketitle

\section{Introduction}
\label{intro}
During decades the geometric Brownian motion has been widely accepted as one of
the most universal models for speculative markets whereas nowadays we
know that many observations of real data are in clear
disagreement with this model~\cite{bouchaudbook}. A possible way out to these inconsistencies is to assume that the volatility is not a constant or even a deterministic function of time or price but a time-depending random variable. Within this approach, there exist the ARCH-GARCH models~\cite{engle82,bolle,baillie,engle-pat} for discrete times and the stochastic volatility (SV) models for continuous times. At late eighties different SV models were presented and all of them proposed a two-dimensional process involving
two variables: the asset price and the volatility~\cite{hull87,scott87,wig87}. These and subsequent works~\cite{Stein,Heston,fouque} were basically aimed to study option pricing. Recently SV models have also been suggested as good candidates to
account for other stylized facts exhibited by the empirical observations in
stocks or indices~\cite{pm,pm2,perello2,Yakov}. Among these facts we want to single out two relevant correlations in time: the volatility autocorrelation and the return-volatility asymmetric correlation.

It is well established that volatility fluctuations have long memory in the
sense that the autocorrelation function decays very slowly with time~\cite{volat1,volat2}. An image of daily data observations is the following: there is a rapid decay governed by a very short time scale (few days) and a fat tail with a time scale at least of the order of several hundred days. However, common SV models result in a single exponential decay implying only one time scale~\cite{hull87,scott87,wig87,pm2}. Therefore we would need to add a second time scale, or even more, to properly reproduce the dynamics of the volatility~\cite{LeBaron}. An alternative approach to model this complex behavior is that of multifractal random walks which assume a power law autocorrelation function that implies
an infinite number of time scales in the evolution of the volatility~\cite{Fisher,muzy}.

On the other hand, the negative return-volatility correlation --also called
asymmetric volatility correlation or leverage-- presents a much shorter time
scale of the order of 10-20 days~\cite{black76,Beckaert,Bouchaud}. Again, the SV
models can easily incorporate the leverage effect with the introduction of an asymmetry parameter. This parameter also induces a bias in the return distribution which in turn
generates the skew observed in many option smiles. Unfortunately, in some SV
models the resulting time scale for the leverage turns out to be comparable to
that of the volatility autocorrelation~\cite{pm,pm2}. Therefore, many SV models seem not to
be able to explain the complex dynamics of the volatility. Some attempts have
been made to harmonize within a single model the different time
scales involved~\cite{masoliver,sircar,vicente}. Even multifractal models
have been recently extended to include leverage effects~\cite{Pochart} although
the resulting models are quite involved and analytical results are scarce and
difficult to obtain.

We have recently addressed this problem and have extended the simple Ornstein-Uhlenbeck (OU) stochastic volatility model by assuming that the mean reverting
level is itself random~\cite{masoliver}. We have thus considered a three-dimensional diffusion process instead of the usual two-dimensional one. With
this model we have been able to obtain two different time scales, one for the
leverage correlation -which, to some extend, coincides with the short time scale
of the volatility autocorrelation- and a second scale which accounts for the
long memory of the volatility process. The model has produced quite satisfactory
analytic results despite the fact of the inherent difficulty in dealing with a
three-dimensional process and also that the plain OU process has small regions where
volatility is not positive definite. Moreover, the OU model, either in its two-dimensional version \cite{perello2} as in the three-dimensional one
\cite{masoliver}, yields a Gaussian probability density function (pdf) for the
stationary distribution of the volatility while several empirical studies have observed that the stationary volatility is far from being
Gaussian~\cite{bouchaudbook,stanley,miciche}.

All these observations have led us to consider a different SV model. Some
authors defend that a good approximation for the
volatility distribution is given by the log-normal distribution or even by the 
inverse Gamma distribution~\cite{bouchaudbook}.
One of our main motivations here is to focus on a model whose volatility is
log-normally distributed. We will thus study the correlated exponential
Ornstein-Uhlenbeck (cexpOU) model insisting on the point of view of its
statistical properties rather than in option pricing as have been done in
previous studies of the model~\cite{fouque}. In this way we will show how the
cexpOU model, despite being a two-dimensional diffusion,
presents a long memory behavior for the volatility with at least two time
scales, the shortest one coinciding with that of leverage. One of the main assets
of the model is that it is still relatively easy to handle and many analytical
results can be obtained.

The paper is divided into six sections. In Section~\ref{anal} we present the
cexpOU model and show its main statistical properties. Section~\ref{correlations} studies the leverage and volatility correlations while Section~\ref{scales} applies the model to actual data thus providing a way of estimating the parameters of the model. The Section~\ref{pdf} studies the probability distribution of the return and conclusions are drawn in Section~\ref{conclusion}.

\section{Analysis}
\label{anal}
As mentioned in the previous section the correlated exponential Ornstein-Uhlenbeck stochastic volatility model has been recently addressed by Fouque {\it et al}~\cite{fouque} in the context of option pricing. Here we present a slight modification of the model which consists in a two-dimensional diffusion process given by the following pair of It\^o stochastic differential equations (SDEs):
\begin{eqnarray}
dX(t)&=&me^{Y(t)}dW_1(t)\nonumber\\
dY(t)&=&-\alpha Y(t)dt+kdW_2(t),
\label{2d}
\end{eqnarray}
where $X(t)$ is the undrifted log-price or zero-mean return defined as
$$
dX=\frac{dS}{S}-\left\langle\frac{dS}{S}\right\rangle,
$$
where $S(t)$ is a financial price or the value of an index. The parameters
$\alpha$, $m$, and $k$ appearing in Eq.~(\ref{2d}) are positive and nonrandom
quantities and $dW_i(t)=\xi_i(t)dt$ ($i=1,2$) are correlated Wiener processes,
{\it i.e.}, $\xi_i(t)$ are zero-mean Gaussian white noise processes with cross
correlations given by
\begin{equation}
\left\langle\xi_i(t)\xi_j(t')\right\rangle=\rho_{ij}\delta(t-t'),
\label{rho}
\end{equation}
where $\rho_{ii}=1$, $\rho_{ij}=\rho$ $(i\neq j, -1\leq\rho\leq 1)$. In terms of
the proces $Y(t)$ the volatility is given by
\begin{equation}
\sigma(t)=me^{Y(t)}.
\label{sigma}
\end{equation}

For the rest of the paper we will assume that at time $t$ the process $Y(t)$ (and hence
the volatility) has reached the stationary state. From Eq.~(\ref{2d}) we see
that the stationary expression for $Y(t)$ is
\begin{equation}
Y(t)=k\int^{t}_{-\infty}e^{-\alpha(t-t')}dW_2(t').
\label{y}
\end{equation}
This expression along with Eq.~(\ref{sigma}) allow us to get the main features
of the stationary volatility. We summarize these features as follows.

The average value, second moment and autocorrelation of the stationary
$\sigma(t)$ are
\begin{equation}
\left\langle\sigma\right\rangle=me^{\beta/2},\qquad \left\langle\sigma^2\right
\rangle=m^2e^{2\beta},
\label{average}
\end{equation}
and
\begin{equation}
\left\langle\sigma(t)\sigma(t+\tau)\right\rangle=	m^2\exp\left\{\beta(1+e^
{-\alpha\tau})\right\},
\label{sigmacorr}
\end{equation}
where
\begin{equation}
\beta=\frac{k^2}{2\alpha}
\label{beta}
\end{equation}
is the stationary variance of $Y(t)$, {\it i.e.},
$\left\langle Y^2(t)\right\rangle=\beta^2$.

The conditional probability density function (pdf) for the volatility is
\begin{equation}
p(\sigma,t|\sigma_0,t_0)=\frac{1}{\sigma\sqrt{2\pi\beta(1-e^{-2\alpha(t-t_0)})}}
\exp\left\{-\frac{[\ln(\sigma/m)-e^{-\alpha(t-t_0)}\ln(\sigma_0/m)]^2}{2\beta(1-
e^{-2\alpha(t-t_0)})}\right\},
\label{sigmapdf}
\end{equation}
which can be derived taking into account that $\sigma(t)$ is an exponential
Ornstein-Uhlenbeck process. The stationary probability density thus reads
\begin{equation}
p_{st}(\sigma)=\frac{1}{\sigma\sqrt{2\pi\beta}}\exp\left\{-\ln^2(\sigma/m)/2
\beta\right\}.
\label{statpdf}
\end{equation}

\begin{figure}
\epsfig{file=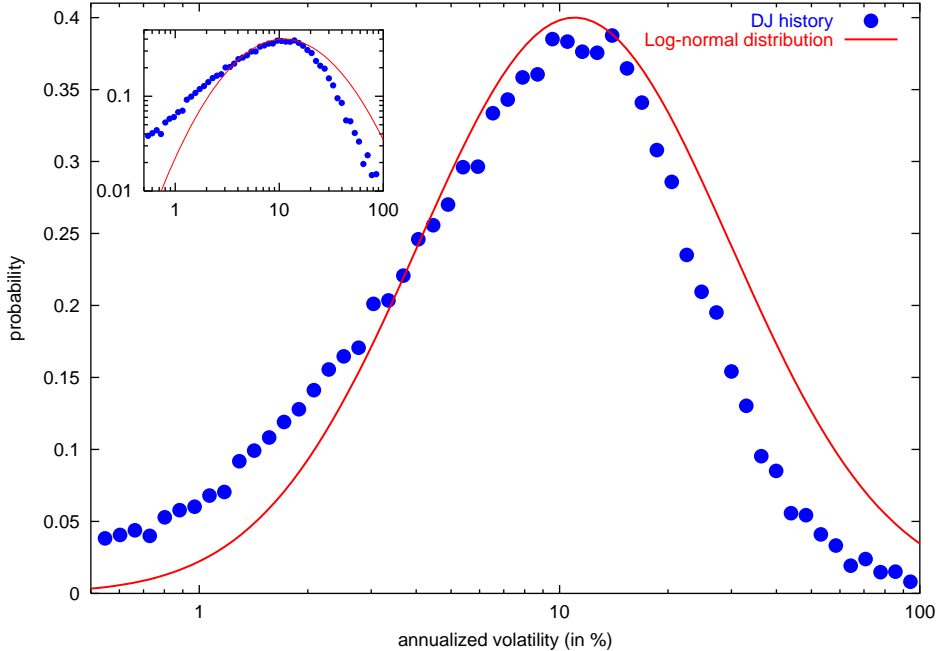}
\caption{Representation of the daily volatility of the DJIA from
1900 to 2004 with a fit using the log-normal distribution given by Eq.~(\ref{statpdf}). We
assume $\sigma$ to be the absolute value of return and a biassed discrepancy is
clearly observed. The inset shows the same representation in a log-log scale in which the discrepancy is enhanced.}
\label{volhistlog}
\end{figure}
\begin{figure}
\epsfig{file=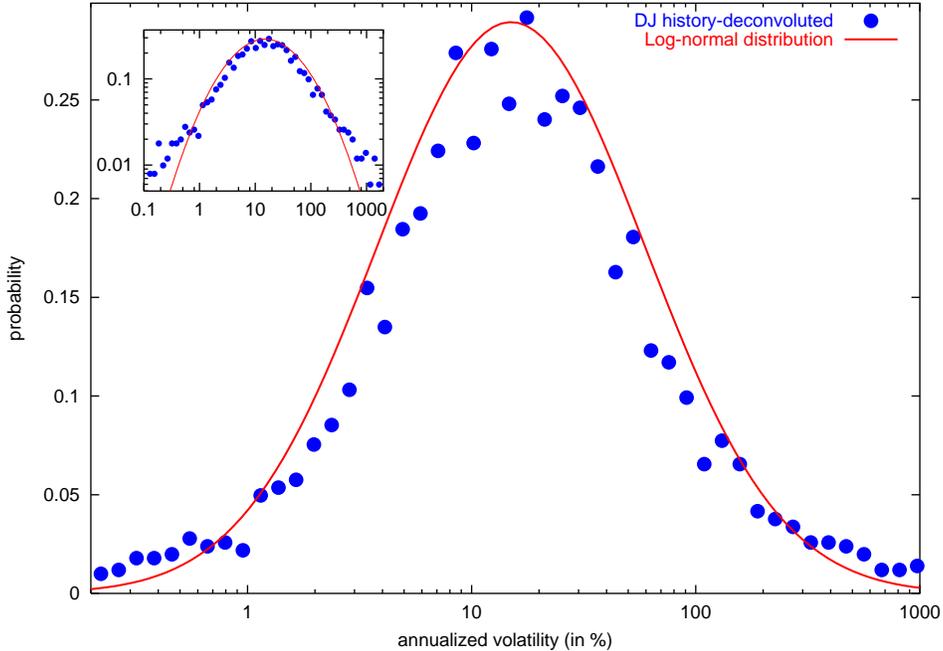}
\caption{Representation of the daily volatility of the DJIA from
1900 to 2004 with a fit using the log-normal distribution given by Eq.~(\ref{statpdf}).
The empirical volatility is obtained using Eq.~(\ref{convolute}) which filters the biass and better fits the log-normal distribution. The inset shows the same representation in a log-log scale.}
\label{volhist}
\end{figure}
\begin{figure}
\epsfig{file=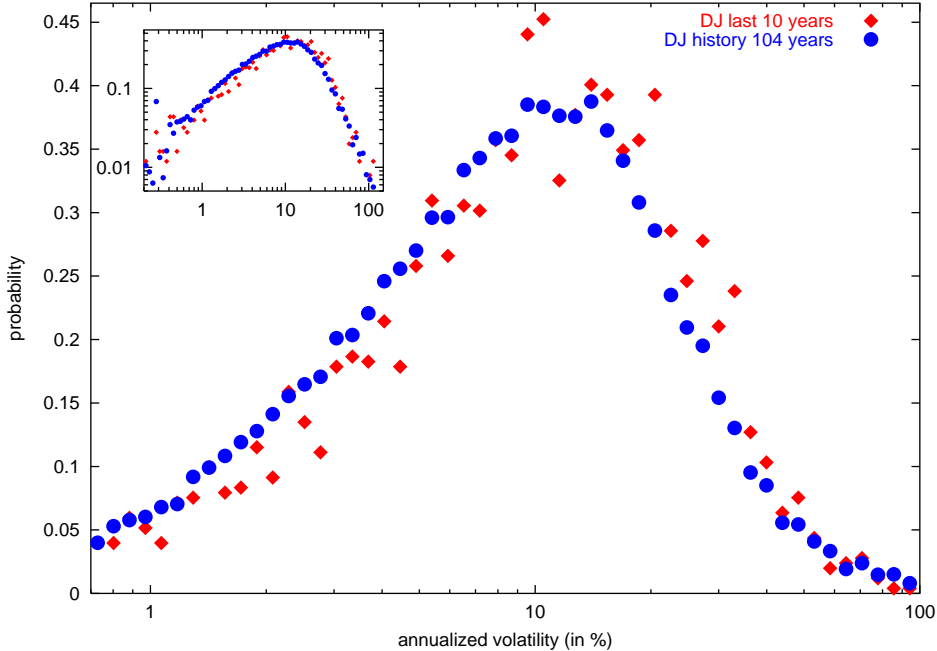}
\caption{Representation of the empirical histogram for the absolute value of
returns estimating the volatility of the DJIA using different data sets. Circles
represent the complete history of 104 years while diamonds represent the last 10
years. The reasonable agreement between the two data sets supports the assumption of stationarity. The inset shows the same representation in a log-log scale.}
\label{volstationarylog}
\end{figure}

We finish this section by confronting the volatility model statistics with daily data of the Dow Jones Industrial Average (DJIA) from 1900 to 2004. The study is limited to sustain the cexpOU model as a good candidate for describing the stationary volatility and we refer the reader to the literature for a more complete analysis 
(see for instance~\cite{stanley,bouchaudbook,miciche}). We first observe that from actual data we only know the time series of prices and the value of $\sigma$ is not directly observed. In practice, one usually takes as an approximate measure of the (instantaneous) volatility the quantity
\begin{equation}
\sigma(t)\approx\frac{\left|X(t+\Delta t)-X(t)\right|}{\sqrt{\Delta t}},
\label{realvolat}
\end{equation}
where $X(t)$ is the zero-mean return. Indeed, since $dX=\sigma dW$ then as $\Delta t\rightarrow 0$:
$$
\left|X(t+\Delta t)-X(t)\right|\approx\sigma(t)|\Delta W(t)|
$$
and this yields Eq.~(\ref{realvolat}) after assuming that $\Delta W(t)^2\rightarrow \Delta t$ as $\Delta t\rightarrow 0$, but this is only true in mean square sense~\cite{gardiner}. Let us check this with empirical data. Figure~\ref{volhistlog} shows the empirical histogram of the stationary volatility in annualized units~\cite{footnote1} using Eq.~(\ref{realvolat}) and we see there a bias in real data which seems to be inconsistent with the log-normal distribution. Hence, the above assumptions may not be correct. An alternative breakout is to deconvolute the product between
$\sigma(t)$ and the absolute value of the Wiener process \cite{footnote1b} (see~\cite{serva} for a more sophisticated method). Thus from the empirical data we obtain the daily return time series $\Delta X=X(t+\Delta t)-X(t)$ and, jointly with a simulation of $|\Delta W(t)|$, we can derive a different estimation of the volatility from:
\begin{equation}
\sigma(t)\approx\frac{\left|X(t+\Delta t)-X(t)\right|}{\left|\Delta W(t)\right|}.
\label{convolute}
\end{equation}
The result is noiser but the biass disappears as we clearly see in
Fig.~\ref{volhist}. We thus get a more reasonable agreement with the theoretical log-normal distribution assumed by our model, a fact that is also confirmed by more complete 
studies~\cite{stanley,bouchaudbook}. On the other hand, Miccich\`e et al.~\cite{miciche} have argued that the log-normal is unable to reproduce the most extreme volatility values and propose a distribution with power law tails. Nevertheless, this regime involves values of sigma greater than 200\% and smaller 1\% (in annualized units) which may be more relevant for high frequency data than for the present case of daily data.

In Fig.~\ref{volstationarylog}, we also plot the empirical histogram of the stationary volatility (with volatility estimated following Eq.~(\ref{realvolat})) taking two different data sets, one embracing the whole period of 104 years and a second one taking the last 10 years of data. The almost complete agreement of the two histograms seems to confirm the assumption of stationarity which is implicit in all the analysis. In any case a more positive conclusion on the stationary assumption would require further tests and analysis~\cite{stanley,bouchaudbook}.

\section{The main correlations}
\label{correlations}

We will now present a deeper look, within the cexpOU model, on the two crucial
correlation mentioned above: the leverage and the autocorrelation of the
volatility. We start by the asymmetric volatility correlation or leverage.

\subsection{Leverage}
\label{lever}

The leverage correlation is difined by
\begin{equation}
{\cal L}(\tau)=\frac{\left\langle dX(t)dX(t+\tau)^2\right\rangle}{\left\langle
dX(t)^2\right\rangle^2},
\label{leveragedef}
\end{equation}
and it is a measure of the correlation between the variations of return and
volatility. Although the anticorrelation between volatility and return has been
known for long~\cite{black76,Beckaert}, it has not been quantitatively studied
until very recently when Bouchaud \textit{et al.}~\cite{Bouchaud}, using a large
ammount of daily returns for both share prices and market indices, showed that
\begin{equation}
{\cal L}(\tau)=\cases{-A e^{-b\tau}, &if $\tau>0$;\cr
0, &if $\tau<0$;}
\label{bouchaudlev}
\end{equation}
$(A,b>0)$. That is, there is an exponentially decaying anticorrelation between
future volatility and past returns changes. No correlation is found between past
volatility and future price changes thus providing a sort of causality to the
leverage effect.

\begin{figure}
\epsfig{file=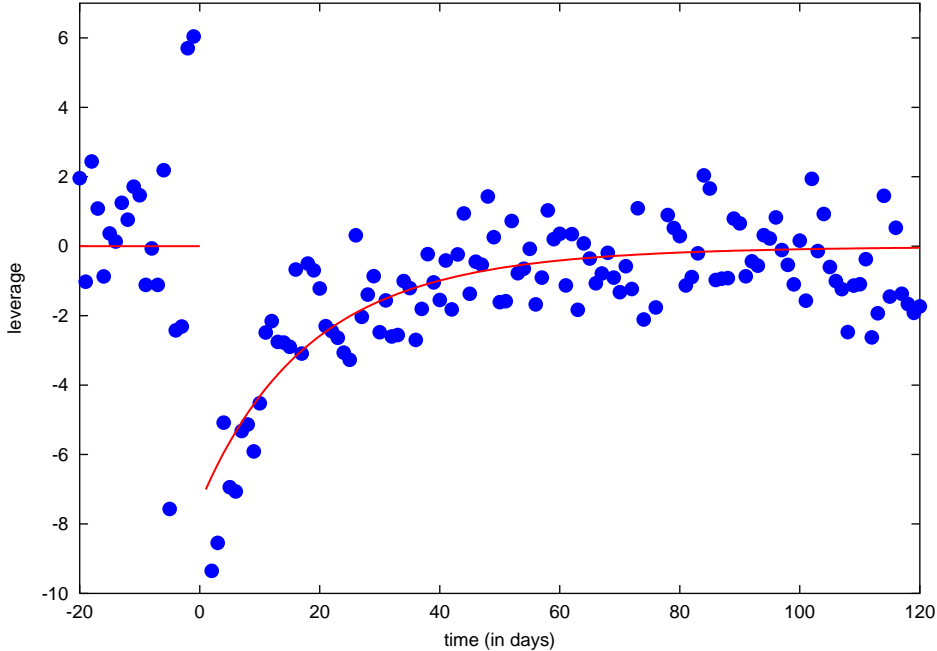}
\caption{The solid line represents the leverage correlation as given by 
Eq.~(\ref{leveragefin}) for the DJIA from 1900 to 2004. Dots represent the empirical data.}
\label{leverage}
\end{figure}

We will now reproduce Eq.~(\ref{bouchaudlev}) using the cexpOU model described
above. From Eq.~(\ref{2d}) and taking into account Eq.~(\ref{sigma}) we have
\begin{equation}
\langle dX(t)dX(t+\tau)^2\rangle=\langle \sigma(t)dW_1(t)\sigma(t+\tau)^2dW_1(t+
\tau)^2\rangle.
\end{equation}
it follows from the It\^o convention that if $\tau>0$ then $dW_1(t+\tau)$ is
uncorrelated
with the rest of stochastic variables. Thus, taking into account that
$\langle dW_1(t+\tau)^2\rangle=dt$,
we have
$$
\langle dX(t)dX(t+\tau)^2\rangle=\langle \sigma(t)dW_1(t)\sigma(t+\tau)^2\rangle
dt
\qquad (\tau>0).
$$
On the other hand when $\tau\leq 0$ the Wiener process $dW_1(t)$ is uncorrelated
with the remaining variables.
Hence, taking into account that $\langle dW_1(t)\rangle=0$, we obtain
$$
\langle dX(t)dX(t+\tau)^2\rangle=0 \qquad (\tau\leq 0).
$$
Therefore,
\begin{equation}
\langle dX(t)dX(t+\tau)^2\rangle=
\langle \sigma(t)\sigma(t+\tau)^2dW_1(t)\rangle H(\tau)dt,
\label{cor1}
\end{equation}
where $H(\tau)$ is the Heaviside step function. In Refs.~\cite{masoliver,pm}, we
have shown that
\begin{equation}
\langle \sigma(t)\sigma(t+\tau)^2dW_1(t)\rangle=2\rho ke^{-\alpha\tau}\langle
\sigma(t)\sigma(t+\tau)^2\rangle dt.
\label{appA1}
\end{equation}
From Eqs.~(\ref{sigmapdf})-(\ref{statpdf}) we get
\begin{equation}
	\langle\sigma(t)\sigma(t+\tau)^2\rangle=m^3\exp\{2\beta(e^{-\alpha\tau}+
	5/4)\},
\label{appA2}
\end{equation}
and
\begin{equation}
	\langle\sigma(t)^2\rangle=m^2e^{2\beta}.
\label{appA3}
\end{equation}
Finally, substituting Eqs.~(\ref{cor1})-(\ref{appA3}) into
Eq.~(\ref{leveragedef}) yields
\begin{equation}
	{\cal L}(\tau)=(2\rho k/m)\exp\left\{-\alpha\tau+2\beta(e^{-\alpha\tau}-
	3/4)\right\}H(\tau),
\label{leveragefin}
\end{equation}
which can have the form given by Eq.~(\ref{bouchaudlev}) although what one has
to identify with $A$ and $b$ will depend on the different scales involved. We
will discuss this issue in the next section. In Fig. \ref{leverage} we plot the leverage
correlation (\ref{leveragefin}) along with empirical data provided by the DJIA.

\subsection{Volatility autocorrelation}
\label{autocor}

\begin{figure}
\epsfig{file=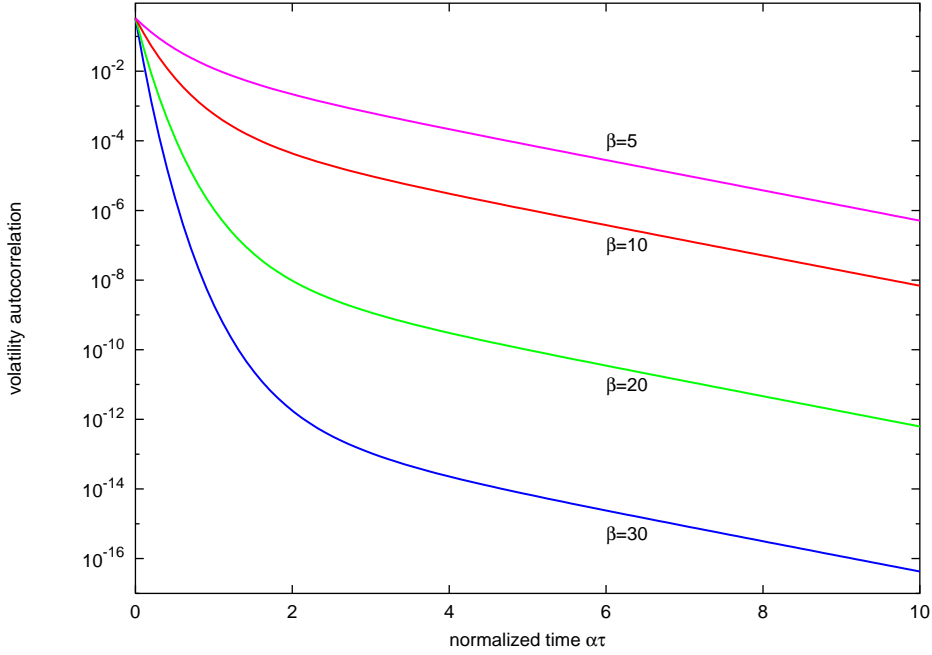}
\caption{We plot in $C(\tau)$ given by Eq.~(\ref{volcorexpand}) as a function
of $\alpha\tau$ for different values of $\beta=k^2/2\alpha$ in a semi log-scale. The figure clearly shows the existence of two asymptotic time scales separated by a sum of multiple
time scales. This effect is enhanced in bigger values of $\beta$.}
\label{prova}
\end{figure}

For the cexpOU model the autocorrelation of the volatility has been already
given by Eq.~(\ref{sigmacorr}). However, as mentioned at the end of Sec.
\ref{anal}, the values of $\sigma(t)$ are not accessible from actual financial
data and we have to use the instantaneous volatility in one of its possible
versions. For this reason, as a measure of the empirical volatility
autocorrelation, we take the function
$C(\tau)=\mbox{Corr}\left[dX(t)^2,dX(t+\tau)^2\right]$, that is,
\begin{equation}
C(\tau)=\frac{\langle dX(t)^2dX(t+\tau)^2\rangle-\langle dX(t)^2\rangle^2}{
\langle dX(t)^4\rangle-\langle dX(t)^2\rangle^2}.
\label{volcor1}
\end{equation}
Again in Ref.~\cite{masoliver}, we have shown that this correlation function can
be written as
\begin{equation}
C(\tau)=\frac{\langle\sigma(t)^2\sigma(t+\tau)^2\rangle-\langle\sigma(t)^2
\rangle^2}{3\langle\sigma^4\rangle-\langle\sigma(t)^2\rangle^2}.
\label{volcor2}
\end{equation}
The averages appearing on the right hand side of this equation can be evaluated
using
Eqs.~(\ref{sigmapdf})-(\ref{statpdf}). The final result reads (see Fig.
\ref{prova})
\begin{equation}
C(\tau)=\frac{\exp[4\beta e^{-\alpha\tau}]-1}{3e^{4\beta}-1}.
\label{volcorfin}
\end{equation}

\section{Determination of parameters and multiple time scales}
\label{scales}

In the cexpOU model, there are four parameters to be determined: $\alpha$,
$\beta$ (or $k$), $\rho$ and $m$. We estimate these parameters as follows.
Firstly, from the long time behavior of the volatility autocorrelation 
(see Eq.~(\ref{volcorlong}) and Fig. \ref{volauto} below) we can determine the value of $\alpha$. Furthermore, from the short time behavior of $C(\tau)$ we can estimate $k$ and hence $\beta$ ({\it cf.} Eq.~(\ref{volcorshort}) anf Fig. \ref{volauto} below). After knowing $\beta$ the value of the normal level $m$ can be obtained through the return variance, that is (see Eqs.~(\ref{average}) and (\ref{realvolat})),
$$
\left\langle \Delta X^2\right\rangle\approx m^2e^{2\beta}\Delta t,
$$
where $\Delta t=1$ day. Finally, the correlation coefficient $\rho$ is estimated
from the value of the leverage correlation, ${\cal L}(\tau)$, as
$\tau\rightarrow 0^+$, that is (see Eq.~(\ref{leveragefin}) and Fig. \ref{leverage}),
$$
{\cal L}(0^+)=\frac{2\rho k}{m}e^{\beta/2}.
$$

For the DJIA and using daily data from January 1, 1900 to June 14, 2004 (a total
of 28,540 data points) the estimated parameters are
\begin{equation}
\alpha=(1.82\pm 0.03)\times 10^{-3} \mbox{ days}^{-1},\qquad
k^2=(1.4\pm 0.2)\times 10^{-2} \mbox{ days}^{-1},
\label{par1}
\end{equation}
which implies $\beta=3.8\pm 0.6$. Finally
\begin{equation}
m=(1.5\pm 0,4)\times 10^{-3} \mbox{ days}^{-1/2} \qquad\mbox{and}\qquad \rho=-
0.4.
\label{par2}
\end{equation}
Note that the value of $m$ corresponds to an annual normal level of volatility
of $(2.4\pm 0.7) \%$.

\subsection{Time scales}

Once determined the relevant parameters we can address the crucial question of
how many time scales have our model. Recall that the empirical volatility needs,
at least, two time scales for daily data~\cite{LeBaron}. In contrast with other two-dimensional stochastic volatility models~\cite{scott87,wig87,hull87,Stein,Heston,Yakov,perello2}, the
cexpOU model has this feature. Indeed, we can write Eq.~(\ref{volcorfin}) in the
form
\begin{equation}
C(\tau)=\frac{1}{3e^{4\beta}-1}\sum_{n=1}^{\infty}\frac{(4\beta)^n}{n!}	e^{-n
\alpha\tau},	\label{volcorexpand}
\end{equation}
which indicates that there are infinite time scales. In Fig. \ref{prova} we plot $\ln C(\tau)$, with $C(\tau)$ given by Eq.~(\ref{volcorexpand}), as a function of $\alpha\tau$ for different values of the parameter $\beta=k^2/2\alpha$. We there observe two relevant time scales as limiting cases of an infinite set of scales (see below). Moreover, the bigger $\beta$ is, the more distant the two time scales are (see Eq.~(\ref{beta2}) and Fig.~\ref{prova}). Had we we plotted other two-dimensional SV 
models~\cite{scott87,wig87,hull87,Stein,Heston,Yakov,perello2} we would have a straight line unable to fit the curve observed in real data.

\begin{figure}
\epsfig{file=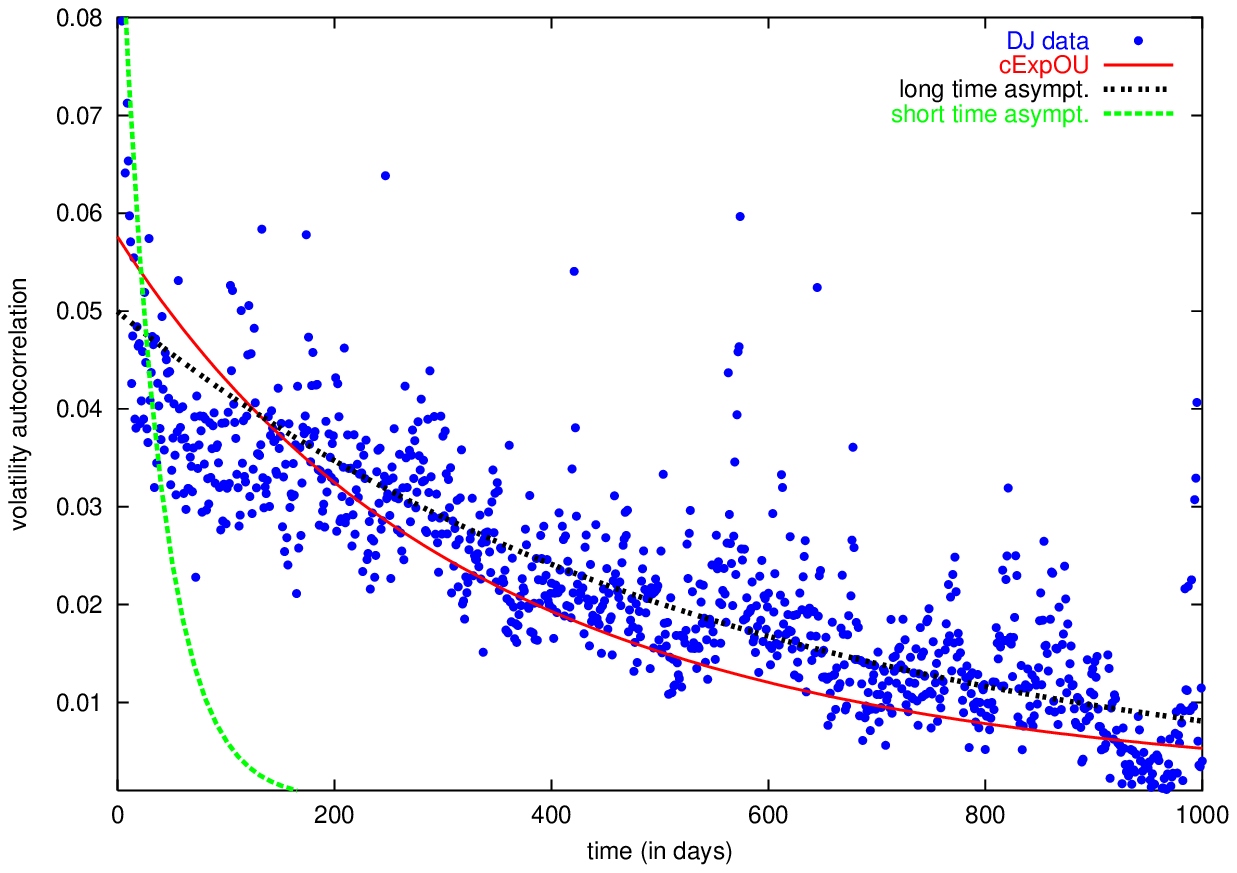}
\caption{The solid line is the volatility autocorrelation as given by
Eq.~(\ref{volcorfin}) and the dashed lines are the asymptotic representations of this equation. Dots represent the empirical data for the DJIA from 1900 to 2004.}
\label{volauto}
\end{figure}
\begin{figure}
\epsfig{file=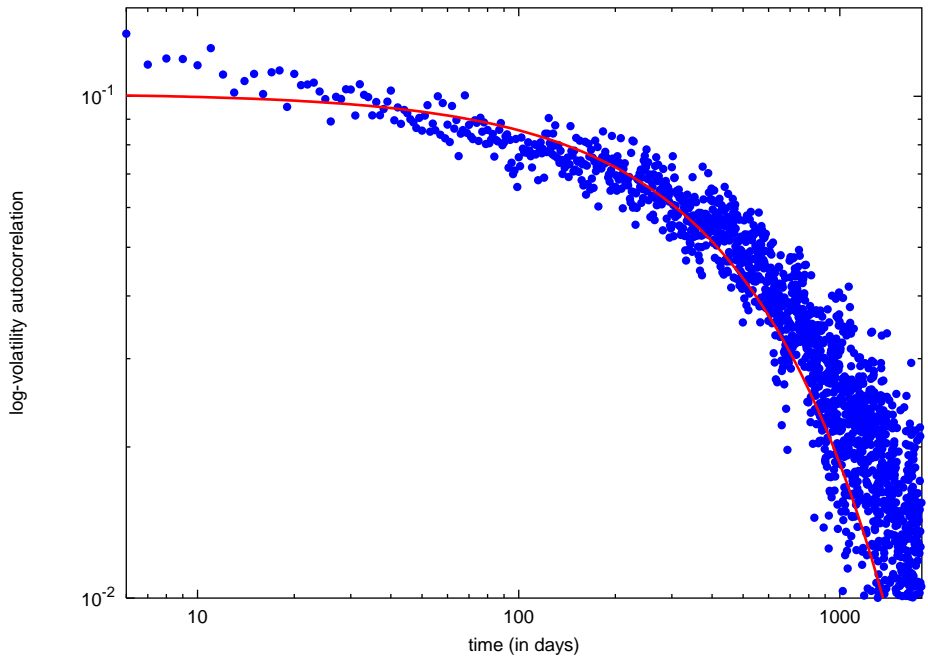}
\caption{Autocorrelation of the process $Y(t)$ in log-log scale. The solid line is
the autocorrelation of $y=\ln(\sigma/m)$ for the DJIA from 1900 to 2004. Dots represent the empirical data.}
\label{volauto-log}
\end{figure}

Let us go deeper in this issue. Suppose first that we are
looking at $C(\tau)$ at long times such that $\alpha\tau\gg 1$. In this case,
neglecting trascendentally small terms of the order of $e^{-2\alpha\tau}$ and
higher in Eq.~(\ref{volcorexpand}), we have
\begin{equation}
C(\tau)\approx\frac{4\beta}{3e^{4\beta}-1}e^{-\alpha\tau}.
\label{volcorlong}
\end{equation}
Thus, the long time behavior of the volatility autocorrelation is governed by
the characteristic time $\alpha^{-1}$ (see Fig.~\ref{volauto}). On the other
hand, for short times such that $\alpha\tau\ll 1$, one can see from Eqs.~(
\ref{beta}) and (\ref{volcorfin}) that
\begin{equation}
C(\tau)=\frac{1}{3e^{4\beta}-1}\left[e^{4\beta-2k^2\tau}-1 \right]+\mbox{O}\left
(\alpha^2\tau^2\right)
\label{volcorshort}
\end{equation}
(see Fig.~\ref{volauto}). In this case the short time behavior of $C(\tau)$ has
the characteristic time $1/2k^2$. In many situations $\alpha^{-1}$ is of the
order of few hundred days (for the DJIA $\alpha^{-1}=549\pm 9$ days) while
$1/2k^2$ is of the order of few weeks (for the DJIA $1/2k^2=35\pm 18$ days)
which is consistent with the separation of time scales just mentioned.
Therefore, within the cexpOU model, the autocorrelation of the volatility
presents, at least, two time scales: the longest one governed by
$\tau_{long}=1/\alpha$ and a shorter one governed by $\tau_{short}=1/2k^2$. We
observe that the dimensionless parameter $\beta$ defined in Eq.~(\ref{beta}) can
be written in terms of $\tau_{long}$ and $\tau_{short}$ as
\begin{equation}
\beta=\frac{\tau_{long}}{4\tau_{short}},
\label{beta2}
\end{equation}
which shows that $\beta$ is a measure of the distance between the long and the
short time scales (see Fig. \ref{prova}).

We can test the soundness of the model by looking at the log-volatility autocorrelation, that is, the autocorrelation of the random process $Y(t)$ (see Eq. (\ref{sigma})). One can easily see from Eq. (\ref{y}) that the cexpOU model has an exponential time decay for the log-volatility:
\begin{equation}
\mbox{Corr}[Y(t+\tau)Y(t)]=(k^2/2m)e^{-\alpha\tau}.
\label{corrY}
\end{equation}
Let us thus see whether the data present the same behavior. To this end we start from the absolute value of the return as an estimator of the daily volatility ({\it cf.} 
Eq. (\ref{realvolat}) with $\Delta t=1$ day) we then take the logarithm of this quantity and evaluate its autocorrelation. We finally compare this empirical correlation with the theoretical one given by Eq. (\ref{corrY}). In Fig. \ref{volauto-log} we show both correlations in log-log scale. We observe that in plotting Eq. (\ref{corrY}) we have used the same parameter values as in Figs. \ref{leverage} and~\ref{volauto}. The agreement between the theoretical model and the empirical data is quite remarkable except, perhaps, at very short time lags where one can observe a small but systematic deviation 
\cite{footnote2}. We can therefore conclude that the cexpOU model provides a fairly satisfactory description of the multiscale behavior of the volatility.

\subsection{The leverage}

The same kind of asymptotic analysis we have performed on the volatility autocorrelation can now be applied to the leverage correlation~(\ref{leveragefin}). Thus for long times, $\alpha\tau\gg 1$, we have
($\tau>0$)
\begin{equation}
{\cal L}(\tau)=\frac{2}{m}\rho ke^{-3\beta/2}\left[e^{-\alpha\tau}+\mbox{O}\left
(e^{-2\alpha\tau}\right)\right],
\label{leverlong}
\end{equation}
which has the form of Eq.~(\ref{bouchaudlev}) with $b=\alpha$ and
\begin{equation}
A=\frac{2}{m}\rho ke^{-3\beta/2}\equiv A_{long}.
\label{along}
\end{equation}
However, as we can realize looking at Eq.~(\ref{beta2}), the value of $\beta$
will be usually large, at least larger than 1. This is precisely the case of
the DJIA where $\beta\simeq 3.8>1$. Consequently $A_{long}$ is exponentially
small and the long-time behavior of ${\cal L}(\tau)$, as expressed by
Eq.~(\ref{leverlong}), turns out to be neglegible and practically undetectable
in
empirical observations. Therefore, the effect of the leverage correlation has to
be sought in the short-time regime $\alpha\tau\ll 1$. In this situation an
expansion of Eq.~(\ref{leveragefin}) yields
\begin{equation}
	{\cal L}(\tau)\simeq \frac{2}{m}\rho ke^{\beta/2} e^{-k^2\tau}\qquad(
	\alpha\tau\ll 1),
	\label{levershort}
\end{equation}
where we have taken into account that $\alpha\ll k^2$. Again
Eq.~(\ref{levershort}) has the form of Eq.~(\ref{bouchaudlev}) with $b=k^2$ and
\begin{equation}
	A=\frac{2}{m}\rho ke^{\beta/2}\equiv A_{short}.
	\label{ashort}
\end{equation}
Note that in those empirical situations in which $\beta\gg 1$:
$$
\frac{A_{long}}{A_{short}}=e^{-2\beta}\ll 1,
$$
thus confirming that the long-time behavior of the leverage correlation is
neglegible and practically undetectable. Therefore, the leverage correlation is
governed by Eq.~(\ref{levershort}) which implies one single time scale given by
$\tau_{lev}=1/k^2$. Note that this time scale is of the same order of magnitude
(a litle longer) than the short time scale of the autocorrelation of the
volatility
({\it cf.} Eq.~(\ref{volcorshort})).

\section{The distribution of returns}
\label{pdf}

We will now address the question of the distribution of returns chracterized by
the return pdf, $p(x,t)$, or, in an alternative and equivalent way, by its
characteristic function $\varphi_X(\omega,t)$.

Let $p_2(x,y,t|y_0)$ be the joint probability density function of the two-dimensional diffusion process $(X(t),Y(t))$ described by the pair of stochastic
differential equations as given in Eq.~(\ref{2d}). This pdf obeys the Fokker-Planck equation
\begin{equation}
\frac{\partial p_2}{\partial t}=\alpha\frac{\partial }{\partial y}(yp_2)+
\frac{1}{2}k^2\frac{\partial^2p_2}{\partial y^2}+
\rho km\frac{\partial^2}{\partial x\partial y}\left(e^yp_2\right)+
\frac{1}{2}m^2e^{2y}\frac{\partial^2p_2}{\partial x^2},
\label{fpe2d}
\end{equation}
with initial condition
\begin{equation}
p_2(x,y,0|y_0)=\delta(x)\delta(y-y_0).
\label{initial2}
\end{equation}
It does not seem to be possible obtaining an exact solution to this problem and
therefore we will search for approximate expressions. There are several possible
strategies to this end depending on the particular values of the parameters of
the model for a given market. Here we will treat the case in which the ``vol of
vol" $k$ is much greater than the normal level of volatility exemplified by $m$.
In other words we will assume that the parameter
\begin{equation}
\lambda=\frac{k}{m}\gg 1
\label{lambda}
\end{equation}
is large. Note that this is the case of the DJIA daily data,
since from Eqs.~(\ref{par1}) and~(\ref{par2}) we see that $\lambda\sim 10^2$.

We define a dimensionless time, $\tau$, and two scaling variables, $u$ and $v$,
by
\begin{equation}
\tau=k^2t,\qquad u=\lambda x,\qquad v=\lambda y.
\label{scaling}
\end{equation}
Then the FPE (\ref{fpe2d}) reads
\begin{equation}
\frac{\partial p_2}{\partial\tau}=\frac{1}{2\beta}\frac{\partial }{\partial v}(
vp_2)+
\frac{1}{2}\lambda^2\frac{\partial^2p_2}{\partial v^2}+
\lambda\rho\frac{\partial^2}{\partial u\partial v}\left(e^{v/\lambda}p_2\right)+
\frac{1}{2}e^{2v/\lambda}\frac{\partial^2p_2}{\partial u^2},
\label{fpescaling0}
\end{equation}
and
\begin{equation}
p_2(u,v,0|v_0)=\delta(u)\delta(v-v_0).
\label{initialscaling0}
\end{equation}
We can remove the dependence on the initial volatility of the joint distribution by performing the average of $p_2(u,v,\tau|v_0)$ over all possible values of $v_0$. If we assume that the initial volatility is in the stationary state, then the unconditional joint density is given by 
$$
p(u,v,\tau)=\int_{-\infty}^{\infty}p_2(u,v,\tau|v_0)p_{st}(v_0)dv_0,
$$
where
$$
p_{st}(v_0)=\frac{1}{\sqrt{2\pi\beta\lambda^2}}e^{-v_0^2/2\beta\lambda^2}.
$$
Hence, $p(u,v,\tau)$ is the solution to the initial-value problem:
\begin{equation}
\frac{\partial p}{\partial\tau}=\frac{1}{2\beta}\frac{\partial }{\partial v}(
vp)+
\frac{1}{2}\lambda^2\frac{\partial^2p}{\partial v^2}+
\lambda\rho\frac{\partial^2}{\partial u\partial v}\left(e^{v/\lambda}p\right)+
\frac{1}{2}e^{2v/\lambda}\frac{\partial^2p}{\partial u^2},
\label{fpescaling}
\end{equation}
and
\begin{equation}
p(u,v,0)=\frac{1}{\sqrt{2\pi\beta\lambda^2}}e^{-v^2/2\beta\lambda^2}\delta(u).
\label{initialscaling}
\end{equation}

We can write a more convenient equation for the characteristic function defined
by
$$
\varphi(\omega_1,\omega_2,\tau)=\int^{\infty}_{-\infty}e^{i\omega_1u}du
\int^{\infty}_{-\infty}e^{i\omega_2v}p(u,v,\tau)dv.
$$
The Fourier transform turns problem (\ref{fpescaling})-(\ref{initialscaling})
into
\begin{eqnarray}
\frac{\partial\varphi}{\partial\tau}=&-&
\frac{1}{2\beta}\omega_2\frac{\partial\varphi}{\partial\omega_2}-
\frac{1}{2}\lambda^2\omega_2^2\varphi(\omega_1,\omega_2,\tau)
\nonumber\\
&-&
\lambda\rho\omega_1\omega_2\varphi\left(\omega_1,\omega_2-i/\lambda,\tau\right)-
\frac{1}{2}\omega_1^2\varphi\left(\omega_1,\omega_2-2i/\lambda,\tau\right),
\label{cf2d}
\end{eqnarray}
and
\begin{equation}
\varphi(\omega_1,\omega_2,0)=e^{-\beta\lambda^2\omega_2^2/2}.
\label{cfinitial}
\end{equation}
Once we know the solution to this problem, the marginal charactistic function of the return $\varphi_X(\omega_1,\tau)$ is obtained through
\begin{equation}
\varphi_X(\omega_1,\tau)=\varphi(\omega_1,\omega_2=0,\tau).
\label{marginalcf0}
\end{equation}
In the Appendix A we show that an approximate expression for $\varphi_X(\omega_1,\tau)$, valid for large values of $\lambda$, is given by 
\begin{equation}
\varphi_X\left(\omega/\lambda,\tau\right)\simeq e^{-\omega^2\tau/2\lambda^2}
\left[1-4i\rho\beta^2a(\tau/2\beta)\frac{\omega^3}{\lambda^3}+4\beta^3b(\tau/2
\beta)\frac{\omega^4}{\lambda^4}+{\rm O}(1/\lambda^5)\right],
\label{cfapprox}
\end{equation}
where the functions $a(z)$ and $b(z)$ are defined in the Appendix A by Eqs. (\ref{a2}) and (\ref{b2}) respectively. 

The inverse Fourier transform of Eq. (\ref{cfapprox}) finally results in an
approximate expression for the pdf of the return which in the original variables
(cf. Eq. (\ref{scaling})) reads
\begin{eqnarray}
p_X(x,t)\simeq\frac{1}{\sqrt{2\pi m^2t}}e^{-x^2/2m^2t}\Biggl[1 &-&
\frac{\rho ka(\alpha t)}{\alpha^{1/2}(2\alpha t)^{3/2}}
H_3\left(-\frac{x}{\sqrt{2m^2t}}\right)\nonumber\\
&+&\frac{k^2b(\alpha t)}{8\alpha(\alpha t)^{2}}
H_4\left(-\frac{x}{\sqrt{2m^2t}}\right)\Biggr],
\label{finalpdf}
\end{eqnarray}
where $H_n(x)$ are Hermite polynomials. Observe that Eq. (\ref{finalpdf}) is an expansion that corrects the Gaussian density:
\begin{equation}
p(x,t)=\frac{1}{\sqrt{2\pi m^2t}}e^{-x^2/2m^2t}.
\label{gaussianpdf}
\end{equation}
This density would correspond to the return pdf if the volatility would
have been a deterministic quantity, {\it i.e.}, if $Y(t)\equiv 0$. We
incidentally note that Eq. (\ref{finalpdf}) constitutes an example of Edgeworth
series, a well stablished approximation procedure in mathematical statistics~
\cite{kendall}.

The deviation from the Gaussian character of the return pdf as shown in Eq.
(\ref{finalpdf}) is also evidenced by the existence of non-zero cumulants of
order higher than two, and specially by the skewness, $\gamma_1$, and the kurtosis
$\gamma_2$. These are related to cumulants by $\gamma_1=\kappa_3/\kappa_2^{3/2}$
and $\gamma_2=\kappa_4/\kappa_2^{2}$. Recall that, knowing the characteristic
function, cumulants can be easily obtained by
$$
\kappa_n=(-i)^n\left.\frac{\partial^n }{\partial\omega^n}\ln[\varphi_X(\omega/
\lambda,\tau)]\right|_{\omega=0}.
$$
Finally from Eq. (\ref{cfapprox}) we get
\begin{equation}
\gamma_1\simeq 6\rho\sqrt{2\beta}\frac{a(\alpha t)}{(\alpha t)^{3/2}},\qquad
\gamma_2\simeq 24\beta\frac{b(\alpha t)}{(\alpha t)^{2}}.
\label{kurtosis}
\end{equation}

\begin{figure}
\epsfig{file=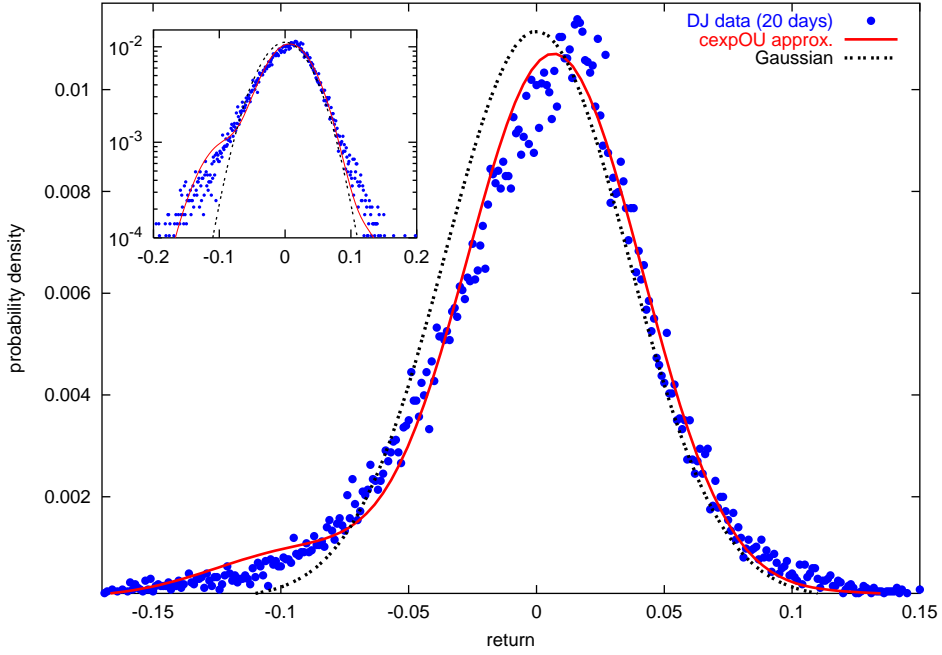}
\caption{The probability distribution of 20 day return for the DJIA. We also
represent the approximate density presented in Eq.~(\ref{finalpdf}) toghether
with the Gaussian pdf given by Eq. (\ref{gaussianpdf}).The inset shows the same representation in a semi-log scale.}
\label{pdfaprox}
\end{figure}
\begin{figure}
\epsfig{file=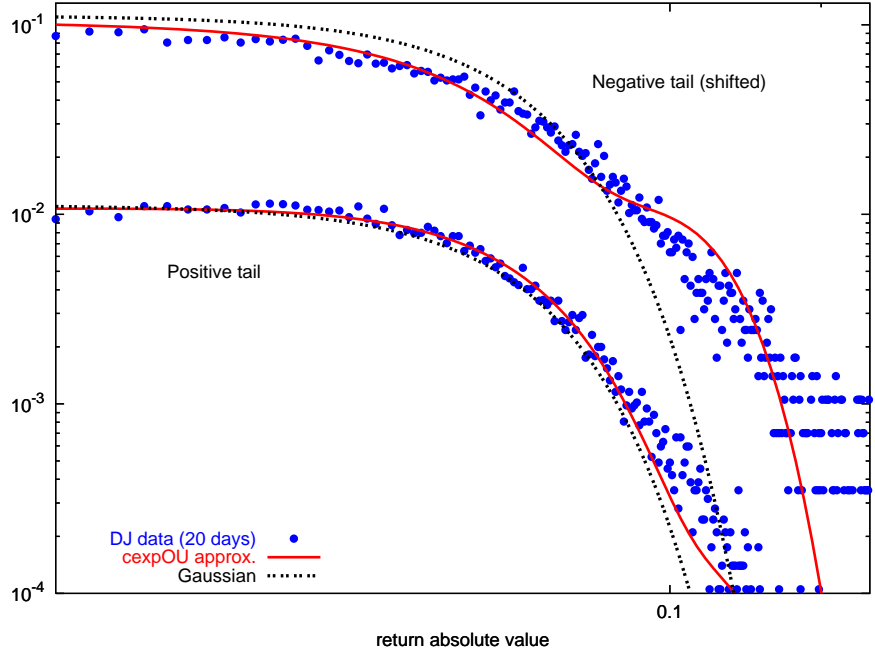}
\caption{The positive and negative tails of the probability distributions
reported in Fig. \ref{pdfaprox}.}
\label{tailaprox}
\end{figure}
In Figs. \ref{pdfaprox} and \ref{tailaprox} we represent the empirical
probability distribution of 20 day returns for the DJIA. We adjust the empirical
distribution with the approximate density given by Eq. (\ref{finalpdf}) which,
as shown in these figures, fits rather well the empirical density in any case
better than the Gaussian pdf (\ref{gaussianpdf}) also represented.

\section{Conclusions}
\label{conclusion}
We have revisited the exponential Ornstein-Uhlenbeck stochastic volatility model. Previous studies on the model were mainly focussed on option pricing somewhat overpassing the description of the underlying asset~\cite{fouquebook}. We have shown that the model explain fairly well some relevant properties of the market. A good feature is that the model has a log-normal stationary probability distribution for the volatility consistent with data. We have found that the model can also explain the observed long-range behavior (with at least two time scales) of the volatility autocorrelation providing also an infinite cascade of time scales. Moreover, we have seen that the values of the parameters that adjust the empirical volatility autocorrelation are also the appropriate ones to describe the short range return-volatility correlation. This consistency has been checked for the particular case of the daily Dow Jones index from 1900 to 2004.

Perhaps one of the main reasons to consider the cexpOU model lies on its simplicity, without the need to resort to fractal phenomena. It still keeps a two-dimensional diffusion formalism which is fully described by only four parameters. Other two-dimensional diffusion models in the literature~\cite{scott87,wig87,hull87,Stein,Heston,Yakov,perello2} cannot describe the multiple time scale behavior and they need to add a third diffusion equation with a second time scale in order to account for the long memory of the volatility~\cite{masoliver,vicente}. The only drawback of the cexpOU is that we do not have an exact solution for its characteristic function in contrast with other SV 
models~\cite{perello2,Yakov}. However, we have also presented an approximate solution starting from the standard Black-Scholes Gaussian distribution. The solution takes into account higher order cumulants and can easily incorporate the effects of skewness and kurtosis. The resultant distribution improves the Gaussian pdf in the statistics of the Dow Jones 20 day returns and may be interesting for option pricing.

\begin{acknowledgments}
This work has been supported in part by Direcci\'on General de Investigaci\'on
under contract No. BFM2003-04574 and by Generalitat de Catalunya under contract
No. 2000 SGR-00023. We thank the participants and organisers of the workshop ``Volatility
of Financial Markets" (Leiden University, 18-29 October 2004) for useful discussions and suggestions.
\end{acknowledgments}

\appendix
\section{The characteristic function of the return}
In order to prove Eq. (\ref{cfapprox}) we will look for an approximate expression of the joint distribution $\varphi(\omega_1,\omega_2,\tau)$ valid for large values of $\lambda$. We first note that the marginal charateristic function of the return 
can be obtained from the joint characteristic function by setting $\omega_2=0$ 
(see Eq. (\ref{marginalcf0})). Therefore,  we will look for a solution to the problem 
(\ref{cf2d})-(\ref{cfinitial}) that for small values of $\omega_2$ takes the form:
\begin{equation}
\varphi(\omega_1,\omega_2,\tau)=\exp\left\{-\left[A(\omega_1,\tau)\omega_2^2
+B(\omega_1,\tau)\omega_2+C(\omega_1,\tau)+{\rm O}(\omega_2^3)\right]
\right\}.
\label{solutioncf2}
\end{equation}
Substituting this into Eq. (\ref{cf2d}) yields
\begin{eqnarray*}
\dot{A}\omega_2^2&+&\dot{B}\omega_2+\dot{C}=-\frac{1}{2\beta}\omega_2(2A\omega_2
+B)+\frac{1}{2}\lambda^2\omega_2^2\\
&+&\lambda\rho\omega_1\omega_2\exp\left\{A\left[2i\omega_2/\lambda+1/\lambda^2
\right]+iB/\lambda\right\}\\
&+& \frac{1}{2}\omega_1^2\exp\left\{4A\left[i\omega_2/\lambda+1/\lambda^2\right]
+2iB/\lambda\right\},
\end{eqnarray*}
where dots represent a time derivative. Expanding the exponentials in powers of $1/\lambda$ up to first order and equating equal powers in $\omega_2$, we obtain an approximate first-order system of ordinary differential equations for $A$, $B$ and $C$ with initial conditions $A(\omega_1,0)=\beta\lambda^2/2$, $B(\omega_1,0)=0$ and $C(\omega_1,0)=0$. The solution to this problem is straightforward and reads
\begin{equation}
A(\omega_1,\tau)\simeq 
\frac{1}{2}\beta\lambda^2e^{-2\gamma(\omega_1)\tau}+
\lambda^2\left[1-e^{-2\gamma(\omega_1)\tau}\right]/4\gamma(\omega_1),
\label{A}
\end{equation}
\begin{eqnarray}
B(\omega_1,\tau)&\simeq&
i\lambda\omega_1^2
\left[\left(2\beta-1/\gamma(\omega_1)\right)e^{-\gamma(\omega_1)\tau}+
\left(1-e^{-\gamma(\omega_1)\tau}\right)/\gamma^2(\omega_1)\right]
\left(1-e^{-\gamma(\omega_1)\tau}\right)/2\gamma(\omega_1)
\nonumber\\
&+&
\lambda\rho\omega_1\left[1-e^{-\gamma(\omega_1)\tau}\right]/\gamma(\omega_1),
\label{B}
\end{eqnarray}
\begin{eqnarray}
C(\omega_1,\tau)&\simeq&\omega_1^2\tau/2+i\rho\omega_1^3
\left[\tau-
\left(1-e^{-\gamma(\omega_1)\tau}\right)/\gamma(\omega_1)\right]/\gamma(\omega_1)
\nonumber\\
&-&
\omega_1^4\left[\tau+
\left(1-e^{-2\gamma(\omega_1)\tau}\right)/2\gamma(\omega_1)
-2\left(1-e^{-\gamma(\omega_1)\tau}\right)/\gamma(\omega_1)\right.
\nonumber \\
&+& \left.
\beta\left(1-e^{-\gamma(\omega_1)\tau}\right)^2\right]/2\gamma^2(\omega_1),
\label{C}
\end{eqnarray}
where
\begin{equation}
\gamma(\omega_1)=\frac{1}{2\beta}-i\rho\omega_1.
\label{gamma}
\end{equation}

From Eqs. (\ref{marginalcf0}) and (\ref{solutioncf2}) we see that the marginal characteristic function of the return reads
\begin{equation}
\varphi_X(\omega_1,\tau)=\exp\left[-C(\omega_1,\tau)\right].
\label{solutionmarginal1}
\end{equation}
Now the return pdf will be given, in the original variables
(cf. Eq. (\ref{scaling})), by the inverse Fourier transform
\begin{equation}
p_X(x,\tau)=\frac{1}{2\pi}\int_{-\infty}^{\infty}
e^{-i\omega x}\varphi_X\left(\omega/\lambda,\tau\right)d\omega.
\label{marginalpdf1}
\end{equation}
We write $\varphi_X\left(\omega/\lambda,\tau\right)$ in the form
$$
\varphi_X\left(\omega/\lambda,\tau\right)\simeq\exp\left\{-\omega^2\tau/2\lambda
^2-
i\rho\omega^3\xi(\omega/\lambda,\tau)/\lambda^3+
\omega^4\psi(\omega/\lambda,\tau)/\lambda^4\right\},
$$
where
$$
\xi(\omega/\lambda,\tau)=\left[\tau-
\left(1-e^{-\gamma(\omega/\lambda)\tau}\right)/\gamma(\omega/\lambda)
\right]/\gamma(\omega/\lambda),
$$
and
\begin{eqnarray*}
\psi(\omega/\lambda,\tau)=\biggl[\tau+
\biggl(1&-&e^{-2\gamma(\omega/\lambda)\tau}\biggr)/2
\gamma(\omega/\lambda)-
2\biggl(1-e^{-\gamma(\omega/\lambda)\tau}\biggr)/\gamma(\omega/\lambda)
\nonumber \\
&+& 
\beta\biggl(1-e^{-2\gamma(\omega/\lambda)\tau}\biggr)\biggr]
/2\gamma^2(\omega/\lambda)
\end{eqnarray*}
Taking into account the expression for $\gamma(\omega/\lambda)$ given by 
Eq. (\ref{gamma}) and expanding $\xi$ and $\psi$ in powers of $1/\lambda$, we get
\begin{equation}
\varphi_X\left(\omega/\lambda,\tau\right)\simeq e^{-\omega^2\tau/2\lambda^2}
\exp\left\{-4i\rho\beta^2a(\tau/2\beta)\frac{\omega^3}{\lambda^3}+
4\beta^3b(\tau/2\beta)\frac{\omega^4}{\lambda^4}+{\rm O}(1/\lambda^5)\right\}
\label{cfapprox0}
\end{equation}
which after a further expansion can be written as
\begin{equation}
\varphi_X\left(\omega/\lambda,\tau\right)\simeq e^{-\omega^2\tau/2\lambda^2}
\left[1-4i\rho\beta^2a(\tau/2\beta)\frac{\omega^3}{\lambda^3}+4\beta^3b(\tau/2
\beta)\frac{\omega^4}{\lambda^4}+{\rm O}(1/\lambda^5)\right],
\label{cfapprox2}
\end{equation}
where
\begin{equation}
a(z)=z-\left(1-e^{-z}\right),
\label{a2}
\end{equation}
and
\begin{equation}
b(z)=(1+2\rho^2)z+2\rho^2\left[ze^{-z}-2\left(1-e^{-z}\right)\right]-
\left(1-e^{-z}\right).
\label{b2}
\end{equation}


\begin{thebibliography}{000}
\bibitem{footnote1} In all figures we represent the volatility in annualized units, that is, we represent the quantity $100\sqrt{252}\sigma$ where $\sigma$ is the daily volatility. Note that we assume $252$ trading days per year and that the factor 100 is added in order to quantify the magnitude in a percent ratio.
\end{thebibliography}
\end{document}